\documentclass[preprint]{aa} 

\usepackage{epsfig}
\usepackage{graphicx}
\usepackage{txfonts}

\begin{document}

\title{Closed-loop focal plane wavefront control with the SCExAO instrument}

\author{
  Frantz Martinache\inst{1}, 
  Nemanja Jovanovic\inst{2,3}, 
  Olivier Guyon\inst{2,4,5}.
}

\institute{
Laboratoire Lagrange, Université Côte d’Azur, 
Observatoire de la Côte d’Azur, CNRS, Parc Valrose, 
Bât. H. FIZEAU, 06108 Nice, France \email{frantz.martinache@oca.eu}
\and 
National Astronomical Observatory of Japan, 
Subaru Telescope, 650 North A’Ohoku Place, Hilo, HI, 96720, USA
\and
Department of Physics and Astronomy, 
Macquarie University, Sydney, NSW 2109, Australia
\and
Steward Observatory, University of Arizona, Tucson, AZ, 85721, USA
\and
College of Optical Sciences, University of Arizona, Tucson, AZ 85721, USA
}

\date{Accepted for publication on April 29, 2016}

\abstract
{}
{This article describes the implementation of a focal plane based wavefront
  control loop on the high-contrast imaging instrument SCExAO (Subaru
  Coronagraphic Extreme Adaptive Optics). The sensor relies on the Fourier
  analysis of conventional focal-plane images acquired after an asymmetric mask
  is introduced in the pupil of the instrument.}
{This absolute sensor is used here in a closed-loop to compensate the
  non-common path errors that normally affects any imaging system relying on an
  upstream adaptive optics system.This specific implementation was used to
  control low order modes corresponding to eight zernike modes (from focus to
  spherical).}
{ This loop was successfully run on-sky at the Subaru Telescope and is used to
  offset the SCExAO deformable mirror shape used as a zero-point by the
  high-order wavefront sensor. The paper precises the range of errors this
  wavefront sensing approach can operate within and explores the impact of
  saturation of the data and how it can be bypassed, at a cost in performance.}
{Beyond this application, because of its low hardware impact, APF-WFS can
  easily be ported in a wide variety of wavefront sensing contexts, for ground-
  as well space-borne telescopes, and for telescope pupils that can be
  continuous, segmented or even sparse. The technique is powerful because it
  measures the wavefront where it really matters, at the level of the science
  detector.}


\titlerunning{Closed-loop focal plane wavefront control with SCExAO}

\maketitle


\section{Introduction}
\label{sec:intro}

Several approaches to high contrast imaging have now clearly demonstrated the
power of focal-plane based image analysis. Most prominently, non redundant
aperture masking (NRM) interferometry \citep{2000PASP..112..555T}, relying on
interferometric calibration tricks in the focal plane has led to high contrast
detections (of the order of 1000:1) in a regime of angular separation
(typically between 0.5 and a few $\lambda/D$) that is still unmatched in
practice by techniques like coronagraphy \citep{2012ApJ...745....5K,
  2015Natur.527..342S}.  As the generation of extreme adaptive optics (XAO)
instruments is coming on-line, more advanced wavefront control schemes
developed in the context of space-borne coronagraphy like speckle nulling
\citep{2006ApJ...638..488B} or the general framework of electric field
conjugation \citep{2009SPIE.7440E..0DG} are being ported on-sky
\citep{2014PASP..126..565M, 2013SPIE.8864E..0KC}.  Nevertheless, it remains
remarkable that such a venerable approach (the original masking idea by Fizeau
was indeed first tested in the 1870s), has remained relevant for well over a
century. This is really a tribute to the deep understanding that interferometry
has brought to the process of image formation.

It was more recently shown that the same self-calibrating tricks used in
masking interferometry, could in fact be applied to regular (i.e. unmasked)
images, assuming AO-correction with residual wavefront errors $\leq 1$
radian RMS.  The notion of closure-phase \citep{1958MNRAS.118..276J}, was
indeed generalized and shown to be a special case of a wider family of
self-calibrating observables coined kernel-phases \citep{2010ApJ...724..464M},
since they form the basis for the null-space (or kernel) of a linear operator.
This generalization also opened the way for a focal-plane based wavefront
sensing approach, relying this time on the eigen-phases of the same linear
operator. While this problem is generally degenerate, one way to break this
degeneracy proved to be simple, and involved masking a small but non negligible
fraction of the pupil to introduce some level of asymmetry.  The principles of
this asymmetric pupil Fourier wavefront sensor (APF-WFS) were described by
\citet{2013PASP..125..422M}, and exploited by \citet{2014MNRAS.440..125P} to
show how it could be used, for instance, to cophase a segmented mirror. 
This paper further expands on the possible applications of this wavefront
sensor, as it is now implemented as part of the SCExAO instrument
\citep{2015PASP..127..890J}, to compensate for a non-common path error unseen
by its upstream pyramid wavefront sensor.

\section{Implementation of closed-loop wavefront control}

\subsection{Theoretical principle of APF-WFS}
\label{sec:theory}

The APF-WFS method relies on the analysis of the Fourier properties of an
AO-corrected image acquired after an asymmetric hard-stop mask has been placed
in the pupil. The SCExAO instrument is equiped with two such masks with the
asymmetric feature at distinct position angles, so that every part of the
instrumental pupil can be accounted for. 
A rotation wheel, located in a plane conjugated with the pupil of the
instrument (cf. Figure 3 of \citet{2015PASP..127..890J}) makes it possible to
move the masks in and out of the beam as required by the observer.

\begin{figure}
\includegraphics[width=\columnwidth]{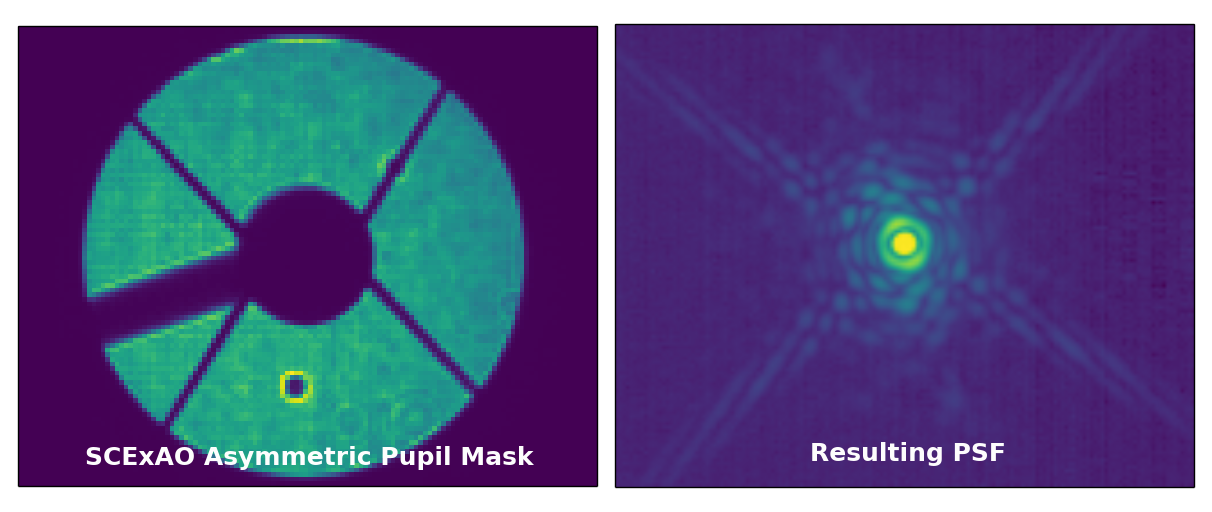}
\caption{Images of the pupil (left) and the focal plane (right) acquired by the
  SCExAO internal science camera. In addition to the four Subaru telescope
  spiders, the thick arm visible on the left hand side of the pupil introduces
  the asymmetry required for the wavefront sensing technique. The thick dot
  visible in the bottom pupil quadrant is induced by a dead actuator on the DM.
  In the focal plane, the presence of this asymmetry results in an additional
  set of diffraction spikes along a direction perpendicular to that of the arm
  and a lumpier first diffraction ring.}
\label{f:1}
\end{figure}

Figure \ref{f:1} shows an image of one of the asymmetric masks in the pupil and
the point spread function (PSF) it produces. Combined with the use of the
2000-element deformable mirror (DM), this simple alteration of the pupil is a
powerful tool used to control the low-order aberrations of the instrument PSF.
All images featured in this paper were acquired using a H-band filter, centered
on wavelength 1.65 $\mu$m and with an effective bandwidth 0.3 $\mu$m. The pixel
scale of the internal science camera is 12.1 mas per pixel, which for this
wavelength, provides a sampling better than Nyquist.

It was shown that, in the low-aberration regime, typical of what is left over
after a first layer of AO correction is applied, the phase $\Phi$ measured in
the Fourier transform of an image $\mathbf{I}$ and the instrumental pupil phase
$\varphi$ are linearly related. 
On the internal calibration source, unaffected by atmospheric turbulence, the
Strehl of images used in this study (such as during the calibration) is
typically of the order of 80\%. On-sky, since the results featured here were
acquired before the XAO loop is closed, the Strehl is significantly lower, of
the order of 50 \%, which is a sufficiently good starting point for
approximation to be valid.

The target phase information, associated with the spatial structures of the
observed object $\Phi_0$, is also present in the Fourier plane and simply adds
to this instrumental Fourier phase. When wavefront aberrations are low (below
$\sim$ 1 radian), the classical image-object convolution relation:

\begin{equation}
  \mathbf{I} = \mathbf{O} \ast \mathbf{PSF},
\end{equation}

\noindent
can therefore be reformulated, if one works with the phase part of the Fourier
transform of this image as follows:

\begin{equation}
  \Phi = \Phi_0 + \mathbf{A} \times \varphi,
  \label{eq:lin}
\end{equation}

\noindent
where $\mathbf{A}$ is an operator that describes the way the pupil phase
$\varphi$ propagates into the Fourier-plane.

When observing a point source, for which $\Phi_0 = 0$ (or if the object is
known), this relation can be inverted if one introduces an asymmetry in the
pupil \citep{2013PASP..125..422M}.  A direct focal plane image, with only a
small amount of additional diffraction generated by the pupil asymmetry
(cf. Figure \ref{f:1}), can therefore serve as a wavefront sensor.

\begin{figure}
\includegraphics[width=\columnwidth]{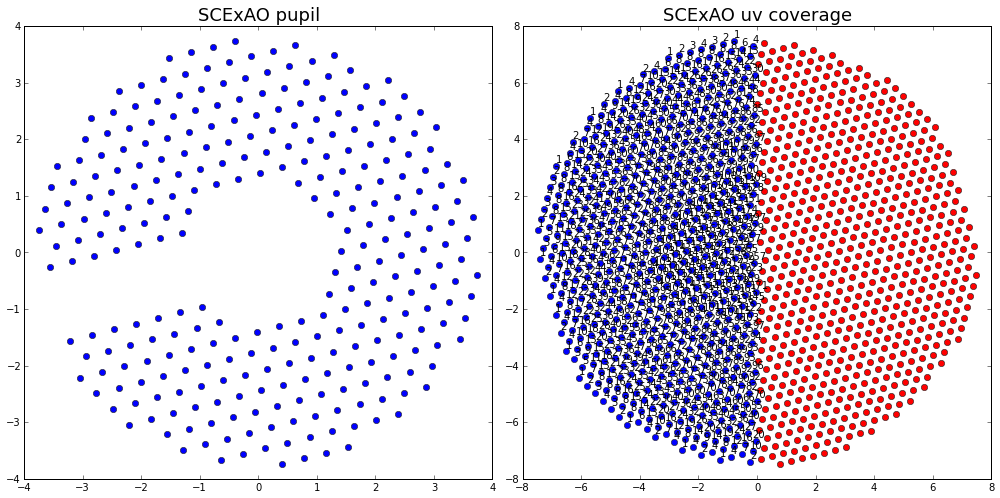}
\caption{ Discrete model of the asymmetric pupil mask used for the calibration
  of the non-common path error in SCExAO. The pupil is discretized into a
  292-element vector that projects onto a set of 675 equivalent interferometric
  baselines (or uv points) in the Fourier domain.  The linear transformation
  that relates the wavefront to the phases measured in the Fourier transform of
  an image is entirely determined from this model. The presence of the
  asymmetry in the pupil ensures that an inverse relation for this phase
  transfer matrix exists.  }
\label{f:2}
\end{figure}

To determine the structure of the operator $\mathbf{A}$, one needs to build a
discrete representation of the instrument pupil - including the asymmetric mask
- following a regular grid with a step such that the sampling density is
reasonably representative of the original pupil.  One then looks at the way
this discrete model projects into equivalent interferometric baselines in the
Fourier plane.  The model currently used on SCExAO is provided in
Figure \ref{f:2}. It reduces the masked pupil to a 292-component vector that
projects onto a 675-element vector in the Fourier domain.  The phase transfer
matrix $\mathbf{A}$ that establishes the mapping between the two spaces ($\Phi
= \mathbf{A} \times \varphi$) is calculated using the \verb+PYSCO+ software,
used for wavefront sensing as well as for kernel-phase data analysis of
diffraction limited images.

The presence of the asymmetry in the pupil ensures that an inverse relation for
this phase transfer matrix exists, and can be used to infer the pupil phase
vector $\varphi$ from the Fourier phase $\Phi$, using the relation:

\begin{equation}
\varphi = \mathbf{A}^{+} \times \Phi
\label{eq:pinv}
\end{equation}

\noindent
where $\mathbf{A}^{+}$ is a Moore-Penrose pseudoinverse of the phase transfer
matrix $\mathbf{A}$, computed after rejecting modes associated to low singular
values. 
The geometry of the asymmetric feature of the mask used for this work
is not the result of an optimization and simply follows the shape used in the
concept paper of \citet{2013PASP..125..422M}. One would expect that a smaller
asymmetry should result in a lower sensitivy but a systematic study of the
sensitivity impact of the geometry of the asymmetry has yet to be done. In the
mean time, the curious reader should check the experimental work of
\citet{2014MNRAS.440..125P}, that shows that, in the case of a segmented
aperture, the technique remains effective, even with a minimum of asymmetry (a
single segment of the aperture) and suddenly breaks, if no asymmetry is present
at all, validating the mathematical model this approach relies on.

\subsection{Integrating a real system}

The case featured in \citep{2013PASP..125..422M} was somewhat idealized,
working on monochromatic images, and with a perfect DM, able to exactly
generate the wavefront correction determined by the analysis.  To deploy this
method on an actual closed-loop system is not as direct and requires us to take
into account the actual properties of the DM, such as the response curve of the
actuators, their influence functions, as well as a careful mapping of these DM
actuators on the instrument pupil \citep{2013PhDT.......417B}.  While possible,
one such model is very prone to errors and its maintenance is demanding,
because of small changes in the internal instrument alignment induced by
temperature drifts or after a telescope slew.

AO systems in operation usually choose to rely on a more pragmatic approach that
encapsulates this kind of model in a transparent manner: individual DM
actuators or groups of actuators (pre-defined modes) are sequentially excited
and the system response is recorded and assembled in a matrix. Filtering of the
noisy modes before inversion (using SVD or similar procedures) leads to the
obtention of a control matrix that can be used to directly multiply an input
vector made of the wavefront sensor raw input data.

\begin{figure}
  \includegraphics[width=\columnwidth]{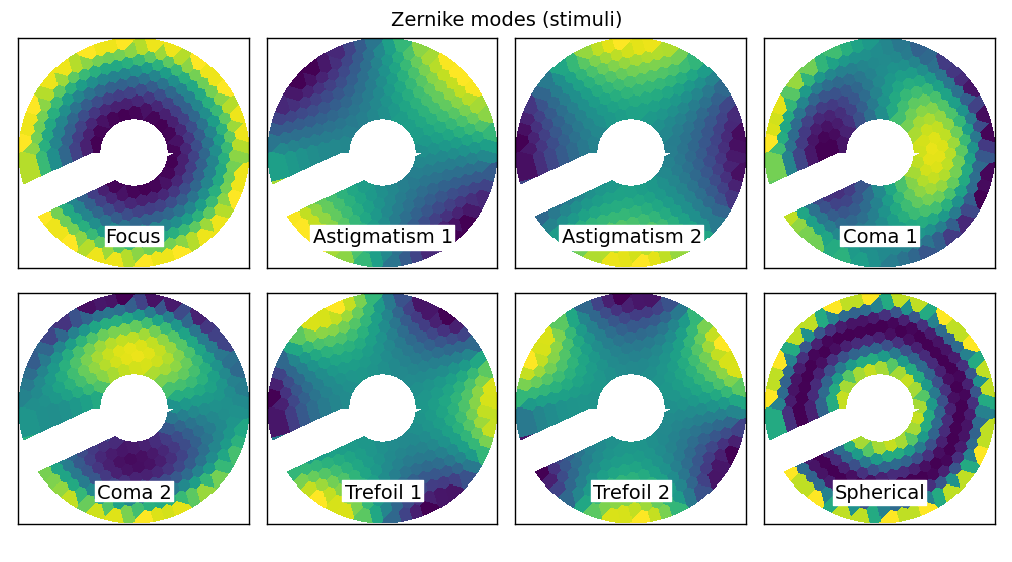}
  \caption{The eight Zernike modes controlled by the SCExAO implementation of
    the APF-WFS. Sorted by Noll index, these modes are (from left to right and
    top to bottom): Z4: focus, Z5: oblique astigmatism, Z6: vertical
    astigmatism, Z7: horizontal coma, Z8: vertical coma, Z9: vertical trefoil,
    Z10: oblique trefoil and Z11: primary spherical.}
  \label{f:3}
\end{figure}

This pragmatic approach is the one that was retained for this implementation of
APF-WFS. The control software was designed to sense and control eight low-order
Zernike modes \citep{1934Phy.....1..689Z} that correspond to classical optical
aberrations: focus, two terms of coma, astigmatism and trefoil, and spherical
aberration.
Figure \ref{f:3} shows how these modes map on the discrete pupil model used
to describe the instrument.

\subsection{Properties of the phase transfer model}

To better appreciate the impact of the phase transfer model, one can look at
the effect of the projection in the Fourier plane and then back in the pupil
plane, using the linear relations of Eq. \ref{eq:lin} and
Eq. \ref{eq:pinv}. The original phase $\varphi$ becomes:

\begin{equation}
\varphi' = \mathbf{A}^{+} \times \mathbf{A} \times \varphi.
\label{eq:recov}
\end{equation}

\begin{figure}
  \includegraphics[width=\columnwidth]{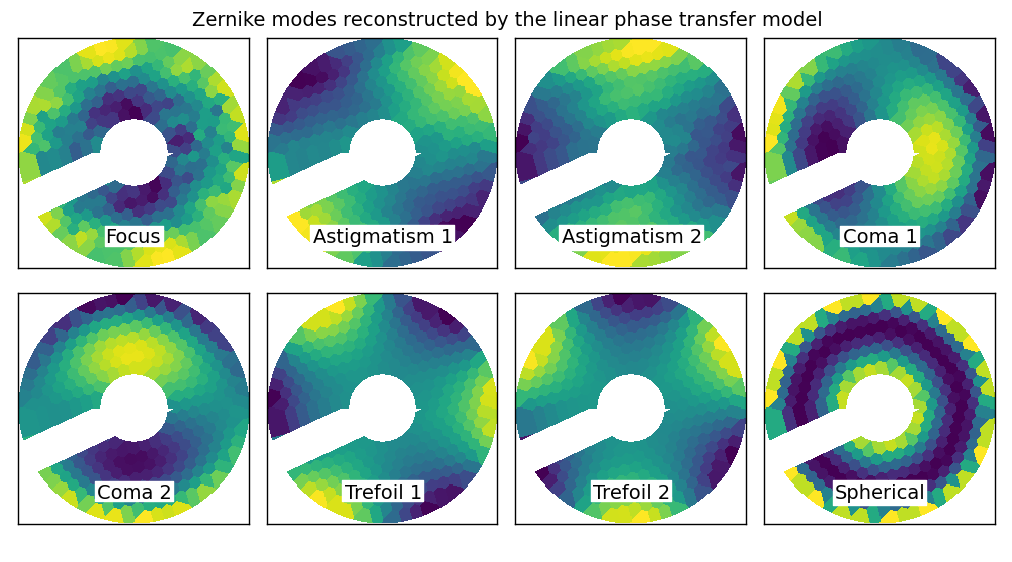}
  \caption{The eight Zernike modes reconstructed by the linear model when 150
    out of the 291 modes are kept in the computation of the pseudo-inverse of
    the phase transfer matrix $\mathbf{A}^{+}$.}
  \label{f:4}
\end{figure}

Depending on the number of modes kept in the determination of the
pseudo-inverse $\mathbf{A}^{+}$ of the phase-tranfer matrix, the reconstruction
goes from perfect (if all modes are preserved) to very partial (if few modes
are preserved). The modes discarded correspond to low singular values, which
for a given level of signal-to-noise in actual data, would result in amplified
noise.  On SCExAO, for the control of these eight low-order modes, 150 out of
the 291 available modes are maintained in the computation of the pseudo
inverse. Under these conditions (cf. Figure \ref{f:4}), the reconstruction
appears satisfactory, and confirms that the technique can indeed be used to
control low-order modes, assuming that the linear model holds (and that
aberrations are small).

\subsection{Calibration}
\label{sec:cal}

The calibration procedure for this implementation of the APF-WFS follows the
linear control framework. After the asymmetric mask has been inserted, one
acquires one image labeled as reference, followed by a sequence of images
acquired after a Zernike mode of appropriate amplitude has been applied to the
DM. Figure \ref{f:4} features one such calibration data-set, acquired with the
focal camera of SCExAO on its internal calibration source (super-continuum
laser) using a standard H-band filter, for a 30 nm RMS deformation of the
DM. Note that this displacement actually translates into a 60 nm wavefront
amplitude modulation (the DM being a reflective system).

\begin{figure}
  \includegraphics[width=\columnwidth]{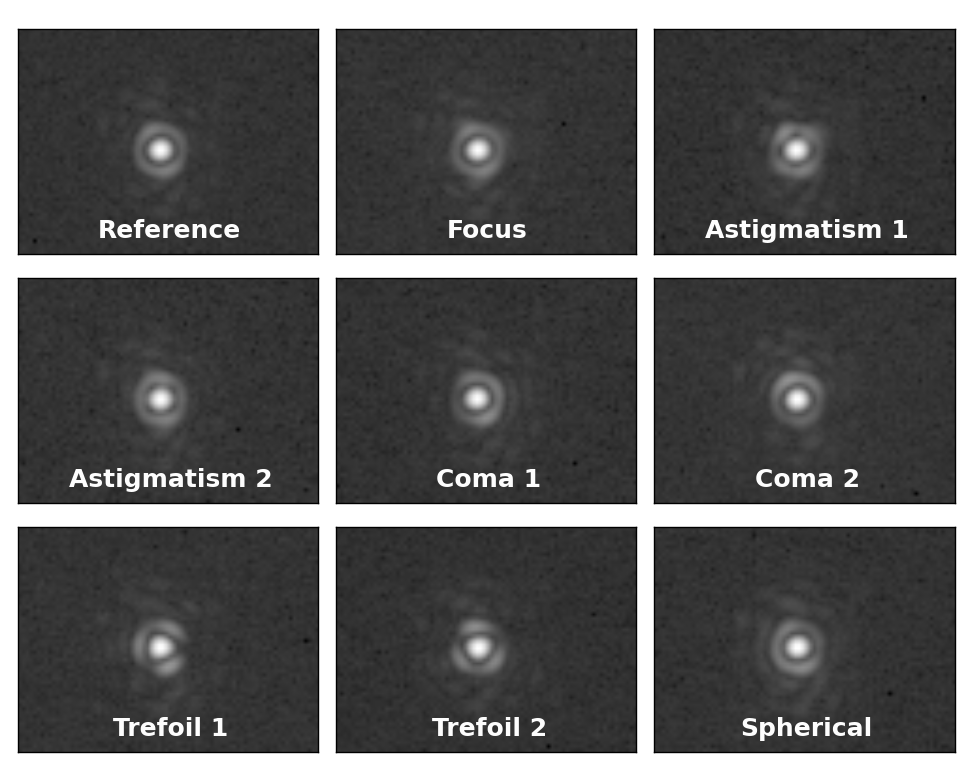}
  \caption{Calibration data for the APF-WFS acquired by the SCExAO science
    camera. Top left: the reference PSF, acquired with the system in its
    starting state. From left to right and top to bottom: the PSF after the
    corresponding Zernike mode has been applied. A non-linear scale is used to
    better show the impact of a 30 nm RMS DM modulation.  }
  \label{f:5}
\end{figure}

Each image is recentered and windowed by a super-gaussian function that filters
out high-spatial frequencies, and reduces the impact of detector readout
noise. It is then Fourier-transformed and the Fourier-phase is extracted
according to the sampling model featured in the right hand side of
Figure \ref{f:2} to populate a vector $\Psi$.

To each Zernike mode, one can therefore associate a Fourier-phase vector
$\Phi_i$ after subtracting the phase $\Psi_{ref}$ measured in the initial or
reference state:

\begin{equation}
  \Phi_i = \Psi_i - \Psi_{ref}.
\end{equation}

\begin{figure}
  \includegraphics[width=\columnwidth]{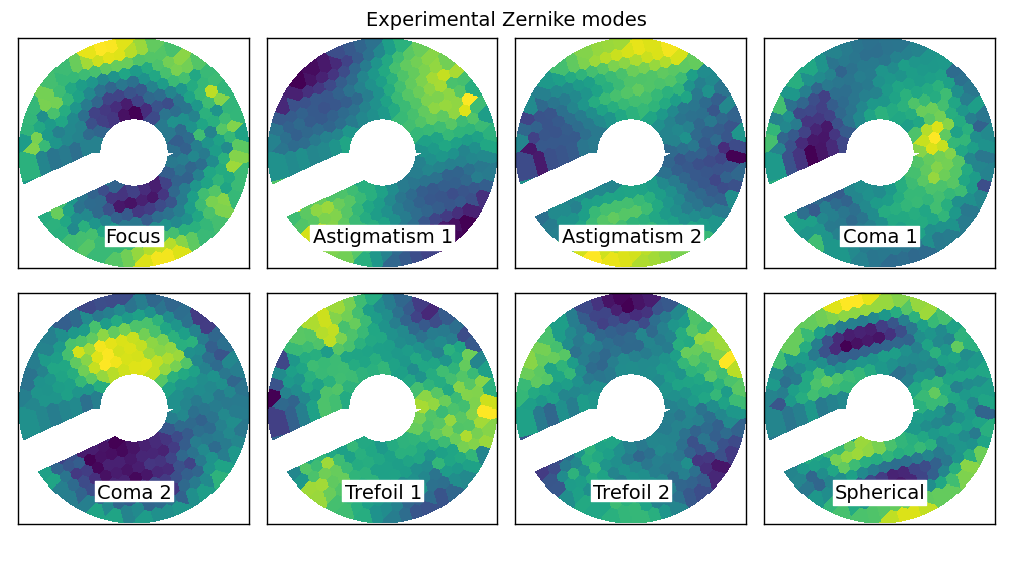}
  \caption{Experimentally recovered Zernike modes. Save for the spherical
    aberration, one will observe that the modes extracted from the analysis of
    the images of Figure \ref{f:5} do reproduce the features expected after
    looking at the theoretical reconstructed modes presented in
    Figure \ref{f:4}.}
  \label{f:6}
\end{figure}

The wavefront associated to this Fourier-phase signature can be recovered using
the pseudo-inverse $\mathbf{A}^{+}$
previously computed and applying Eq. \ref{eq:pinv}.  This wavefront in radians
can in turn be converted into a DM displacement map (in microns) after being
multiplied by the proper $\lambda/4\pi$ scaling factor; where $\lambda$ is the
wavelength expressed in microns and $4\pi$ contains the $\times$2 factor due to
the reflection. Figure \ref{f:6} features an example of experimentally
recovered modes. 
One will observe that the reconstruction from the Fourier analysis of actual
images appears visually satisfactory. Differences in the reconstruction with
the modes plotted in Figure \ref{f:4} can be attributed to imperfections of the
pupil discretization model (cf. Figure 4 of \citet{2013PASP..125..422M}),
combined to practical subtleties like the fact that the dynamical range on the
camera is limited and that classical noises: photon and readout, do apply.

\begin{figure*}
  \includegraphics[width=\linewidth]{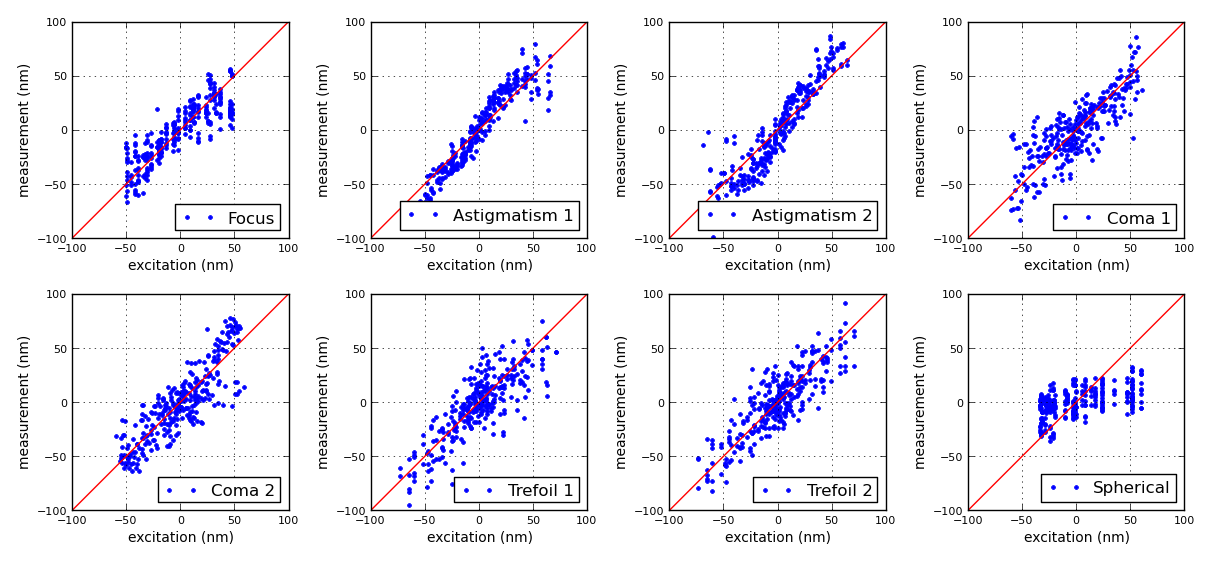}
  \caption{Comparative quality of the modal reconstruction: for each mode, the
    local value of the DM displacement (in nm) for the theoretical Zernike mode
    of predefined amplitude (here 30 nm, labeled as the 'excitation') is
    plotted against its experimental measurement.}
  \label{f:7}
\end{figure*}

To complete this description of the general aspect of the modes with a
quantitative estimate of the quality of the reconstruction, Figure \ref{f:7}
directly compares the experimental reconstruction $E$
to the theoretical Zernike modulation $T$, by plotting the local deduced DM
displacement against its predicted value. 
One can confirm that all modes are not equally reproduced by the analysis, and
that for this experimental setup, the sensor is most sensitive to astigmatism
(for which the correlation coefficient is the strongest) and not particularly
suited to the sensing of spherical aberration. 
Table \ref{tbl:quality} accompanies this figure and provides the value of the
Pearson product moment correlation coefficient for all modes:

\begin{equation}
  r = \frac{\mathrm{cov}(E, T)}{\sigma_{E}\sigma_{T}},
\end{equation}

\begin{table}
\begin{center}
\begin{tabular}{  l  l }
\hline
\bf{Zernike Mode} & \bf{quality} \\
\hline
Z4 (focus)         & 0.855 \\
Z5 (astigmatism 1) & 0.949 \\
Z6 (astigmatism 2) & 0.925 \\
Z7 (coma 1)        & 0.820 \\
Z8 (coma 2)        & 0.862 \\
Z9 (trefoil 1)     & 0.827 \\
Z10 (trefoil 2)    & 0.854 \\
Z11 (spherical)    & 0.522 \\
\hline
\end{tabular}
\caption{Pearson Product moment correlation coefficient of the experimentally
  reconstructed mode $E$ with their theoretical counterpart $T$.}
\label{tbl:quality}
\end{center}
\end{table}

Table \ref{tbl:quality} also shows that with a correlation coefficient
$\sim$0.5, spherical aberration is significantly less well reconstructed than
the other modes that exhibit correlation coefficient $>$0.8.
The specificity of the response to spherical aberration is however not an
intrinsic limit of the sensing approach and can in fact simply be explained by
the 2D geometry of this aberration and how it fits within the footprint of the
Subaru Telescope and its large ($\sim$30\% central obstruction). 
By looking, for instance at Figure \ref{f:3}, one will observe that the donut
shape of the spherical aberration results in a pretty uniform distribution of
the phase, that only varies near the inner and outer edges of the pupil. The
basis of Zernike polynomials is defined for a complete circular aperture:
quoted amplitudes correspond to a given wavefront RMS over the entire circular
aperture. For the spherical aberration, the presence of the central obstruction
naturally filters a lot of the effect of the spherical aberration. 
This is further confirmed after a close examination of scatter plots of
Figure \ref{f:7}: whereas all Zernike stimuli (along the horizontal axis of the
plots) have the same amplitude, one will observe that the resulting range of
local DM displacement (corresponding to the horizontal spread of the data
points) is appreciably shorter for the spherical aberration than it is for the
other modes.

These experimentally obtained pupil-phase modes $\varphi_i$ are stored in a $8
\times 291$ matrix $\mathbf{Z}$, refered to as the response matrix. In
practice, unless the DM registration were to change in a dramatic manner, the
calibration is quite robust: a response matrix acquired using the internal
calibration source and can be very well be used during on-sky observations, if
the filter remains unchanged and if the change of exposure time does not
result in a saturated PSF core (cf. the discussion in Section \ref{sec:disc}).

On SCExAO, the acquisition of this response matrix only takes a few seconds, so
it can easily be repeated if needed after acquisition of a new target. In
practice, it seems a response matrix acquired on the stable internal
calibration source provides the best results.

\subsection{Closed-loop operation}
\label{sec:cloop}

Just like during the calibration, focal-plane images acquired on-sky with the
asymmetric mask are dark-subtracted, recentered and windowed by a
super-gaussian function before being Fourier-transformed. After extraction of
the Fourier-phase, a wavefront is produced and directly projected onto the
basis of modes (without subtracting the reference), to find the coefficients
associated to all eight Zernike components.  If the current wavefront sensor
signal is $\varphi$, the instant Zernike coefficients $(\alpha)$ are the
solution of $\mathbf{Z} \cdot \alpha = \varphi$. The least square solution
$(\hat{\alpha_i})_{i=4}^{11}$ of this system is the solution to:

\begin{equation}
  \mathbf{Z^T} \mathbf{Z} \cdot \hat{\alpha} = \mathbf{Z^T} \varphi
\end{equation}

The solution $\hat{\alpha}$ to this well behaved system of equations is used as
an input for a control loop algorithm. The loop in operation on SCExAO
implements a simple proportional controller, with a gain common to all Zernike
modes with value contained between 0.05 and 0.3, depending on the overall stability of the
wavefront provided by the upstream AO. When looking at the internal calibration
source, one can reliably use the highest gain. Since it is for now only used
for a very short time (typically $\sim15$ seconds), at the time of target
acquisition to flatten the static component of the wavefront, the current
implementation of the algorithm is proving satisfactory.
Once the non-common path error is accounted for, the asymmetric mask is taken
out of the optical path and the system is ready for observing using the full
pupil of the telescope.

\section{Performance}

\subsection{On-sky demonstration}

\begin{figure}
  \includegraphics[width=\columnwidth]{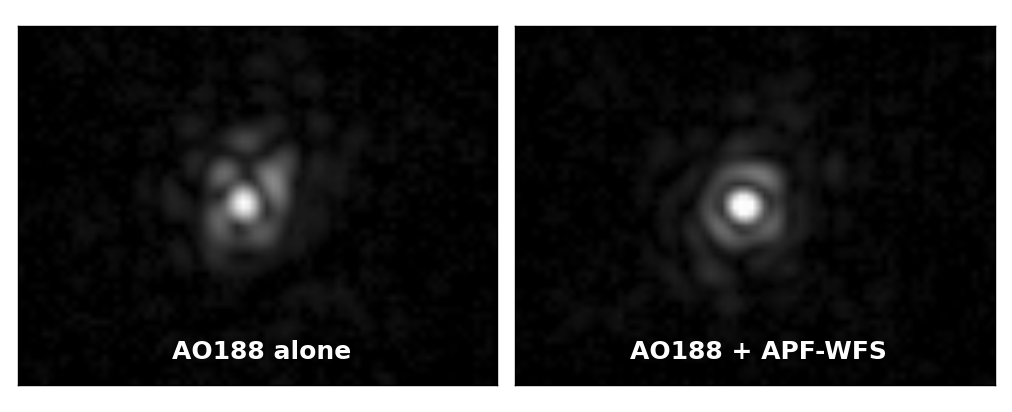}
  \caption{Illustration of the impact of the APF-WFS. Left: 0.5 ms PSF acquired
    by SCExAO's internal science camera after the upstream AO loop has been
    closed. Right: identical exposure acquired 30 seconds after the APF-WFS
    loop has been closed. Despite residual imperfections due to dynamic
    changes, the PSF quality is obviously improved.}
  \label{f:8}
\end{figure}

The technique was successfully deployed and proved to be effective at reducing
the non-common path error during on-sky observations behind Subaru Telescope's
facility AO system AO188 \citep{2010SPIE.7736E.122M}.  Figure \ref{f:8}
illustrates the impact of the approach, with two 500 $\mu$s exposures of the
target (Altair) acquired by SCExAO's internal science camera on UT 2015-10-30.

The first image shows the PSF after the AO188 loop has been closed on the
target: although it features a well defined diffraction core, the PSF clearly
exhibits some static aberrations that can be attributed to the non-common path
error between AO188 and SCExAO's focal plane. The second image shows the PSF
about 30 seconds after the APF-WFS loop has been closed. The gain in Strehl is
low (of the order of 5 \%), but the PSF is improved at where it matters most for
high contrast imaging and no longer features any obvious low-order aberration
signature. Residual inhomogeneity of the first diffraction ring can be
attributed to a combination of instantaneous AO residuals combined with the
effect of the asymmetric arm.

SCExAO's internal science detector is a fast but low-sensitivity detector that
can acquire images at up to 170 Hz full-frame rate whose specifications are
given in \citep{2015PASP..127..890J}. 
APF-WFS seems to exhibit sufficient sensitivity to be used in a fast closed-loop
that could very well track low-order aberrations with frequencies up to a tenth
of the camera frame rate.

At the moment, the goal of the loop is to calibrate the quasi-static non-common
path error. The control software keeps a rolling average of the 20 last
wavefront estimations, and corrects for the average of these estimations at
each iteration, thus filtering vibration-induced fast varying component.
Combined to the acquisition, the (non-optimized) computation of the wavefront
makes the loop run at a frequency of $\sim$8 Hz.

\subsection{Cross-talk}

Zernike polynomials \citep{1934Phy.....1..689Z} form a convenient basis to
describe a wavefront within a circular aperture: designed to form an
orthonormal basis, the first terms of the series happen to correspond to
classical monochromatic optical aberrations like focus, astigmatism or coma.
As previously seen, the presence of a central obstruction in the pupil makes
this basis no longer perferctly orthogonal. Substitutes have been proposed
\citep{1981JOSA...71...75M} in order to accomodate for the presence of this
rather common feature of telescopes, but the asymmetric arm required for the
wavefront sensing (cf. Section \ref{sec:intro}) would also require an
adaptation.

\begin{figure}
  \includegraphics[width=\columnwidth]{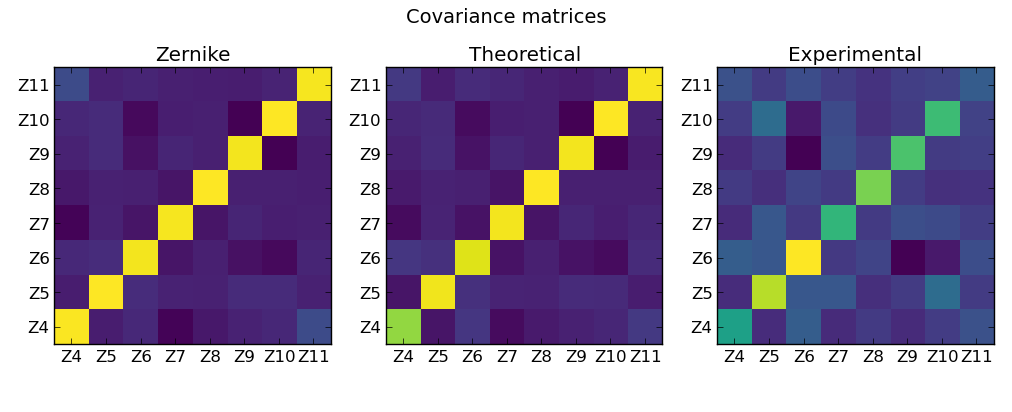}
  \caption{Control modes inner product matrices caracterizing the orthogonality
    properties of the implementation of the APF-WFS described in this paper on
    SCExAO. From left to right: 1. the mostly perfectly diagonal case of the
    Zernike polynomials basis, 2. the seemingly identical inner product matrix
    for the theoretical modes after reconstrucion by the linear model and
    filtering by the SVD, and 3. the inner product matrix for the
    experimentally reconstructed modes. The overall diagonal allure of the
    latter characterizes the sensor as suited to the control of the low order
    Zernike modes. Numerical values for the experimental inner products are
    provided in Table \ref{tbl:dot}.
  }
\label{f:9}
\end{figure}

Instead of trying to specify a new orthogonal basis perfectly adapted to our
case, we have judged more appropriate to stick to the conventional Zernike
basis, and verify {\it a posteriori} how orthogonal the different modes
actually are.  Figure \ref{f:9} does this by plotting the $8 \times 8$
matrix of inner products between the eight control modes
($\mathbf{Z^T} \mathbf{Z}$) for three cases: 
the input Zernike polynomials (given in Figure \ref{f:3}), the theoretical
reconstruction of the linear model with 150 out of the 291 eigen modes kept in
the phase transfer matrix inversion (given in Figure \ref{f:4}) and the
experimentally acquired modes (given in Figure \ref{f:6}).

An orthogonal basis will result in a perfectly diagonal inner-product matrix
whereas non-orthogonality would become manifest with strong non-diagonal
components.

One can therefore observe, looking at the left hand side panel of Figure
\ref{f:9}, that the Zernike modes do form a satisfactory, nearly orthonormal
basis, with a mostly uniform diagonal and a limited amount of cross terms
standing out (except for the case of the 16\% cross-correlation between focus
({\bf Z4}) and spherical ({\bf Z11}).  For the sake of consistency with the
rest of the data presented in the paper, Figure \ref{f:9} also shows in its
central panel, that the modes reconstructed by the linear model reproduce most
of these features, although one can observe $\sim$20\%
degradation of the relative strength of the focus signal. What we observe here
is the effect of the filtering of low singular values in the construction of
the pseudo-inverse $\mathbf{A}^{+}$
as used in Equation \ref{eq:recov}.  With 150 out of the possible 291 modes
kept in the construction of the pseudo-inverse, the inner product matrix for
the experimentally recovered modes is also mostly diagonal.

Table \ref{tbl:dot} provides the numerical values for the experimental inner
products, also graphically represented in the right hand side panel of Figure
\ref{f:9}. Although some of the cross terms are non-neglibible, the terms along
the diagonal still dominate, indicating that a control loop relying on this
calibration dataset will reliably converge toward a state that will cancel the
non-common path aberration.

\begin{table}
\begin{center}
\begin{tabular}{  l | r r r r r r r r }
 &\bf{Z4}&\bf{Z5}&\bf{Z6}&\bf{Z7}&\bf{Z8}&\bf{Z9}&\bf{Z10}&\bf{Z11} \\
\hline
\bf{Z4}  &  0.46 & -0.09 &  0.12 & -0.10 & -0.04 & -0.09 & -0.03 &  0.06 \\
\bf{Z5}  & -0.09 &  0.87 &  0.09 &  0.09 & -0.08 & -0.03 &  0.19 & -0.03 \\
\bf{Z6}  &  0.12 &  0.09 &  1.00 & -0.04 &  0.01 & -0.25 & -0.16 &  0.05 \\
\bf{Z7}  & -0.10 &  0.09 & -0.04 &  0.57 & -0.04 &  0.05 &  0.03 & -0.02 \\
\bf{Z8}  & -0.04 & -0.08 &  0.01 & -0.04 &  0.75 & -0.03 & -0.07 & -0.07 \\
\bf{Z9}  & -0.09 & -0.03 & -0.25 &  0.05 & -0.03 &  0.65 & -0.03 & -0.02 \\
\bf{Z10} & -0.03 &  0.19 & -0.16 &  0.03 & -0.07 & -0.03 &  0.61 &  0.00 \\
\bf{Z11} &  0.06 & -0.03 &  0.05 & -0.02 & -0.07 & -0.02 &  0.00 &  0.11
\end{tabular}
\caption{Numerical values for the experimental modes inner-product}
\label{tbl:dot}
\end{center}
\end{table}

\subsection{Range of linear response}
\label{sec:lin}

\begin{figure}
  \includegraphics[width=\columnwidth]{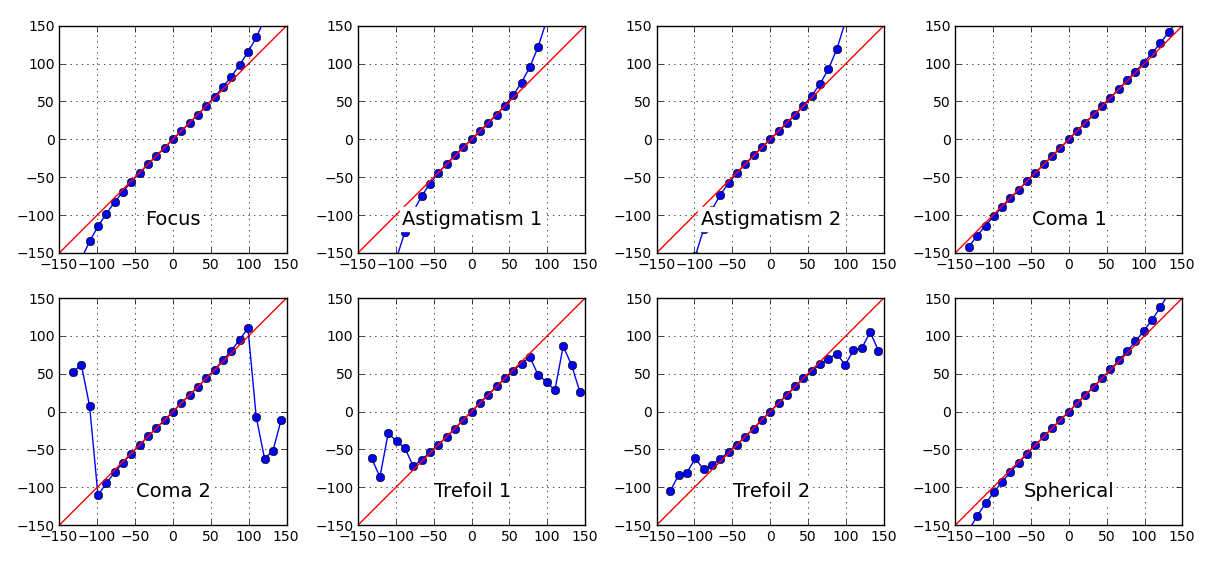}
\caption{Experimental response of the APF-WFS obtained on the SCExAO internal
  (super-continuum) source in the H-band. Each plot features (on the vertical
  axis) the reponse of the sensor to a Zernike mode of RMS amplitude that
  varies over a $\pm$150 nm range (units for both axis are in nm). 
  Nearly linear over the entire range for most modes, the sensor only exhibits
  a significant non-linear behavior for the coma 2, and the two trefoil modes
  when the DM Zernike amplitude is larger than 80-100 nm. Note that this limit
  is on the DM surface, which must be doubled if refering to aberrations on the
  wavefront.}
\label{f:10}
\end{figure}

As reminded in Section \ref{sec:theory}, the APF-WFS relies on the assumption
that an upstream AO correction is provided. The system is expected to deal with
with small residual wavefront errors, and the calibration procedure described
above, typically employs DM modulation amplitudes of $\sim$30-50
nm. To determine the amount of aberration the technique is able to deal with,
we performed a systematic exploration of the response of the sensor to stimuli
of variable amplitude.
The instantaneous response of the sensor is projected onto the basis of modes
following the procedure outlined in Section \ref{sec:cloop}. Figure \ref{f:10}
summarizes the result of this systematic exploration of the response of the
sensor, over a $\pm$150 nm range of DM modulation amplitude.

Although not perfectly linear, for {\bf Z4} (focus), {\bf Z5}, {\bf Z6}
(astigmatism), {\bf Z7} (coma 1) and {\bf Z11} (spherical), the response
remains monotonic over the entire $\pm$150 nm range. For {\bf Z8} (coma 2),
{\bf Z9} and {\bf Z10} (trefoil 1 and 2), the response is only monotonic
over the $\pm$80 nm modulation range beyond which the sensor cannot be used
reliably.

The drastic difference of response between {\bf Z7} and {\bf Z8} which
correspond to the same type of aberration (ie. coma), can be explained by
the azimuth of the asymmetric arm in the pupil stop, oriented such that it
masks out almost entirely one of the two antisymmetric bumps that are
characteristic of this aberration (cf. for instance the top-left most panel of
Figure \ref{f:3}).

A strong non-linearity of the response is experienced when the pupil-phase
peak-to-valley (P2V) wavefront becomes larger than 2$\pi$ (which results in a
phase wrap).
The presence of the asymmetric stop at its current azimuth essentially divides
the P2V by a factor of two in the case of {\bf Z7} (ie. {\bf coma 1}), hence
making the sensor able to handle twice as much coma along the horizontal axis
than along the vertical axis.
Note that the same effect (to a lesser extent) can also be observed when
comparing {\bf Z9} and {\bf Z10}.

We can nevertheless conclude that under normal operating conditions, in the
H-band, the sensor is able to operate linearly as long as the RMS error on
either mode is less than 200 nm on the wavefront.

\section{Wavefront sensing from a saturated PSF}
\label{sec:disc}

A well AO-corrected PSF is a highly contrasted object. The proper
simultaneous sampling of the core of the PSF and its diffraction rings
therefore requires a detector with a large dynamical range, that is rarely
compatible with a fast readout. The SCExAO internal science camera has an
effective dynamical range of $\sim$10000 counts, so that in practice, one has
to choose an exposure time that either gives access to a non-saturated PSF core
(the normal operating mode of the wavefront sensing approach described thus
far) or over-expose the PSF core to better see the fainter diffraction
structures that surround it.

In its general form, the linear model of Eq. \ref{eq:lin} only holds when
working on non-saturated images that otherwise result in a non-translation
invariant PSF. Pixels that are saturated by the bright core of the PSF can be
treated as zeros, so that the effect of saturation can be modeled as a
multiplication by a top-hat function that cuts off anything higher than a
level imposed by the characteristics of the detector. This multiplication
in the image space, results in a convolution by an Airy-like function whose
characteristic size is inversely proportional to the size of the saturated part
of the PSF. 

\begin{figure}
  \includegraphics[width=\columnwidth]{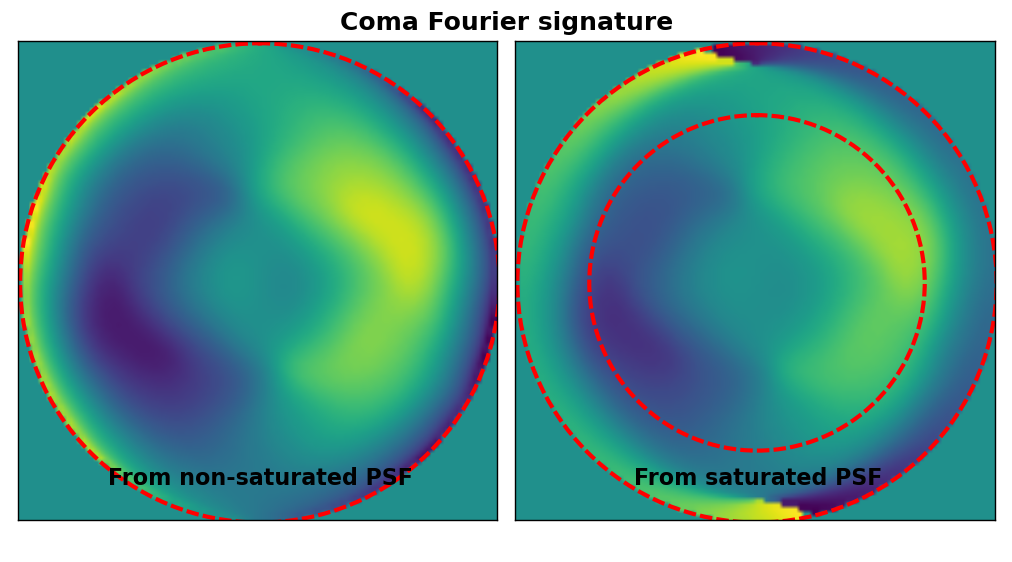}
\caption{Comparison of the uv-phase signature of the same amount of aberration
  (here coma) for a non-saturated PSF (on the left) and a saturated one (on the
  right). The saturation primarily affects the higher spatial frequencies of
  the image, corresponding to the outermost parts of the Fourier plane. In the
  inner region of the saturated case (within the highlighted smaller circle),
  the Fourier-phase signature of the aberration is similar to its non-saturated
  counterpart.}
\label{f:11}
\end{figure}

The effect of this convolution is expected to be most prominent in the
outermost region of the Fourier plane where the phase will experience a change
of sign. Figure \ref{f:11} illustrates this effect by comparing the Fourier-phase
signature of a specific aberration for a non-saturated PSF to that of a mildly
saturated one. One can observe that while the outermost part of the
Fourier-phase is obviously affected by the saturation, the innermost part of
the PSF (highlighted by the smaller dashed circle) does resemble the original
non-saturated case.

\begin{figure}
  \includegraphics[width=\columnwidth]{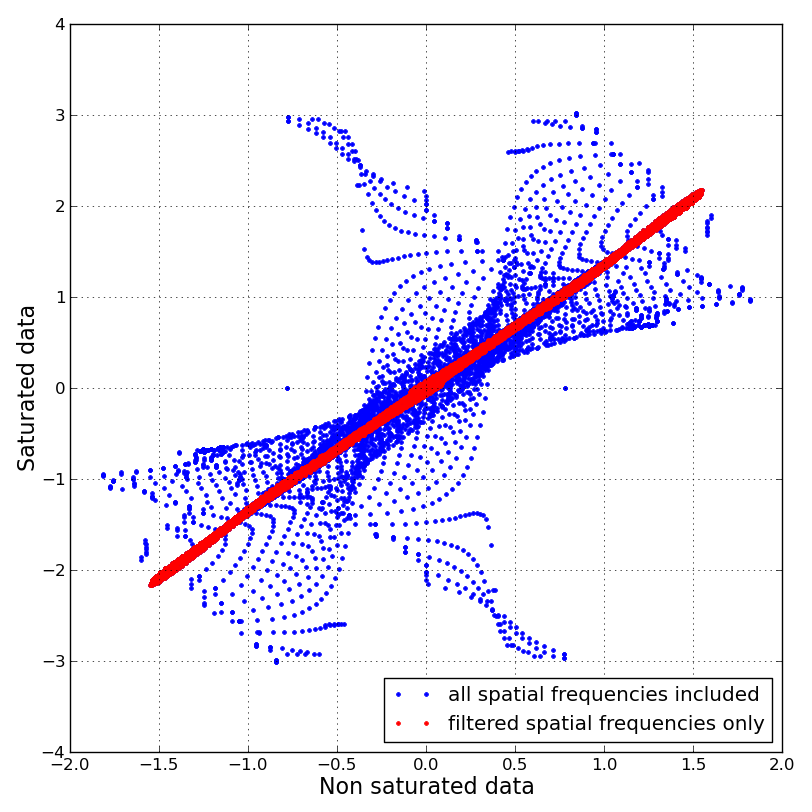}
  \caption{1D comparison of the phase (in radians) extracted from the Fourier
    plane featured in Figure \ref{f:11}, in the non-saturated case (along the
    horizontal axis) and in the saturated case (along the vertical axis). The
    blue points include the data at all spatial frequencies while the red ones
    correspond to the inner part of the Fourier plane only. The strong
    correlation observed in the latter case suggests that some of the wavefront
    information can be recovered from the analysis of mildly saturated data.
}
\label{f:12}
\end{figure}

Figure \ref{f:12} further makes this obvious by representing in a 1D plot the
values of the Fourier-phase of the saturated image against the Fourier-phase of
the non-saturated image. Whereas considered as a whole, the Fourier-phase of
the saturated data appears as non-usable (the blue points are widely
scattered), the inner part of this same saturated data set is strongly
correlated (the red points) with the non-saturated data, suggesting that some
of the wavefront information can be recovered in the saturated case, assuming
that one filters out the information coming from the largest baselines.

To account for this filtering, the Fourier-phase model can be modified, and the
parts of the phase transfer matrix $\mathbf{A}$ can be discarded along with the
parts of the Fourier-phase vector $\Phi$ that are filtered out. The model we
have tested only preserves baselines that are 70\% or less than the longest
baseline in the model, which corresponds to the area inscribed within the inner
circle plotted in the right panel of Figure \ref{f:11}.
Out of the 675 original distinct uv-phase samples, 330 remain with this
configuration, which is still of the order of the number of modes one needs in
order to recover the full theoretical pupil phase information (291 modes). 

\begin{figure}
  \includegraphics[width=\columnwidth]{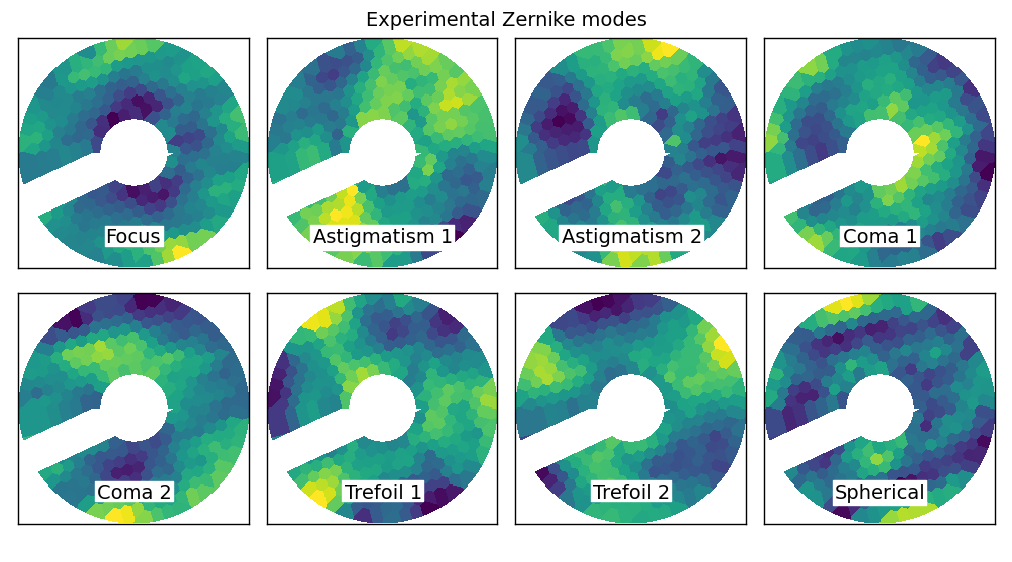}
\caption{Experimentally recovered Zernike modes after discarding the
  phase associated to the longest baselines, affected by saturation
  effects. This new series of experimental modes should be compared to the
  non-saturated case presented in Figure \ref{f:6}.
}
\label{f:13}
\end{figure}

For the computation of the pseudo-inverse $\mathbf{A}^{+}$ of
this new system, 50 modes are kept. With less constraints from the uv-plane,
the Zernike modes are less well reconstructed, but are nevertheless
recognizable, as shown in Figure \ref{f:13}.

\begin{figure}
  \includegraphics[width=\columnwidth]{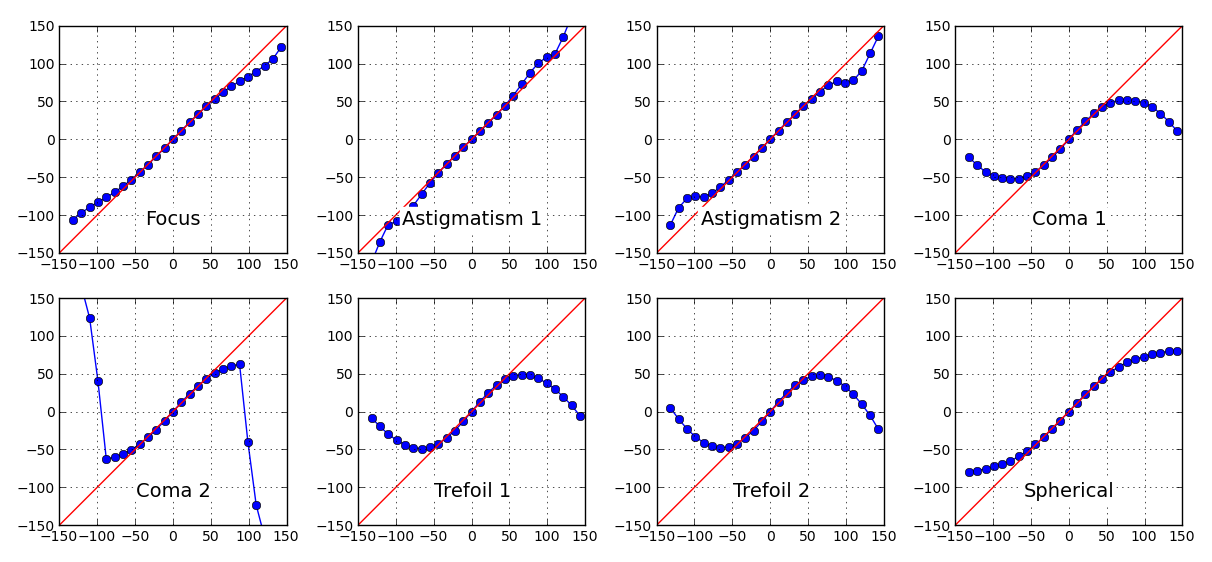}
\caption{Study of the response of the sensor in the saturated case to Zernike
  modes of varying amplitude. This series is to be compared to the case
  presented in Figure \ref{f:10} in the non-saturated case.
}
\label{f:14}
\end{figure}

The calibration procedure introduced in Section \ref{sec:cal} can be repeated
with this new model after which, the APF-WFS is effectively able to operate in
closed-loop from the analysis of saturated data, albeit with lower performance.
To see how this saturation affects the sensor, the study presented in
Section \ref{sec:lin} was repeated in this new operating mode. The outcome of
this study is presented in Figure \ref{f:14}.

In this peculiar saturated mode, the sensor is able to operate linearly over a
limited range of aberrations. Like previously observed (cf. Section
\ref{sec:lin}), it is the modes whose geometry feature localized bumps like
coma of trefoil that first limit the range of aberration that APF-WFS can
account for. APF-WFS can nevertheless operate on images whose core is
saturated, over a 100 nm wavefront RMS range that is roughly one half of what
it can achieve in the non-saturated operating regime: the general APF-WFS
approach is more robust than first expected and can be used despite less than
ideal conditions.

\section{Conclusion}

Following on the conceptual study proposed by \citet{2013PASP..125..422M}, this
paper described the implementation of the asymmetric pupil Fourier wavefront
sensor as one of the wavefront control loops of the SCExAO instrument.
This approach has proven able to repeatedly account for the non-common path
error that affects the instrument after a new telescope pointing and provide an
updated zero-point for the upstream pyramid wavefront sensor currently
implemented inside SCExAO.

A surprisingly simple asymmetric hard stop mask introduced in the pupil of a
diffraction limited imaging instrument is therefore proving to be a powerful
diagnostic tool for the control of the non-common path aberrations. The
reported capture range of the technique is currently limited to a fraction of a
wave (RMS $\sim\lambda/8$).
A combination of filters of decreasing wavelengths would provide a direct way
to tolerate a cruder starting point. We are currently exploring the potential
of an updated algorithm that simultaneously exploits the information sampled at
multiple wavelengths to extend the capture range even more, this time within
the coherence length.  Note that other approaches using combinations of
non-redundant aperture masks \citep{2012OExpr..2029457C, 2014OExpr..2212924C}
also rely on this idea to extend their capture range.

The asymmetry results in slight cosmetic degradations of the PSF. While this
does impact a coronagraphic instrument, it can be tolerated in a general
purpose AO-corrected imaging instrument. 
A very interesting feature of this image-based wavefront sensing approach is
that if multiple sources are available in a given field, the APF-WFS algorithm
can be used on all sources simultaneously. Depending on the complexity of the
field, the same asymmetric mask, combined with the analysis of multiple sources
in one image can be used for multi-reference wavefront sensing, opening the way
to a full 3D reconstruction of the wavefront from the analysis of a single
focal plane image.
This very property can also be put to use on artificially introduced incoherent
replicas of an on-axis PSF, of tunable intensity as described by
\citet{2015ApJ...813L..24J}, thus making the use of the technique compatible
with that of a coronograph that otherwise destroys the interferometric
reference required for a sensible Fourier-analysis of the image as described in
this work.

The use of this wavefront control technique extends well beyond the control of
low-order modes on SCExAO: this paper provides experimental evidence that the
technique is actually effective where the theory predicts it should be. 
In an exposure that simultaneously features an unsaturated PSF core and the
diffraction features at large separation with sufficient SNR, APF-WFS can be
used to control an arbitrary number of modes, as was shown in the concept
paper. APF-WFS can in fact easily be applied in a wide variety of wavefront
sensing contexts, for ground- as well space-borne telescopes, and with a pupil
that can be continuous, segmented or even sparse. APF-WFS is powerful because
it measures the wavefront where it really matters, at the level of the science
detector. Given its low impact on the instrument hardware, it is an option that
should be given some consideration, as part of any high contrast imaging
instrument with wavefront control capability.


\end{document}